\documentclass[conference]{IEEEtran}
\IEEEoverridecommandlockouts
\usepackage{cite}
\usepackage{amsmath,amssymb,amsfonts}
\usepackage{algorithmic}
\usepackage{graphicx}
\usepackage{textcomp}
\usepackage{xcolor}
\usepackage{verbatim}
\usepackage{hyperref}
\usepackage{float}
\def\BibTeX{{\rm B\kern-.05em{\sc i\kern-.025em b}\kern-.08em
    T\kern-.1667em\lower.7ex\hbox{E}\kern-.125emX}}
\begin{document}

\title{Effects of dynamic power electronic load models on power systems analysis using ZIP-E loads \\
\thanks{Thanks to \textit{Réseau de Transport d'Électricité} for partially funding this work.}
}

\author{\IEEEauthorblockN{Gabriel E. Col\'on Reyes, Claire Tomlin}
\IEEEauthorblockA{\textit{Dept. of Elect. Engineering and Comp. Science} \\
\textit{University of California, Berkeley} \\
\{gecolonr, tomlin\}@berkeley.edu}
\and
\IEEEauthorblockN{Reid Dye}
\IEEEauthorblockA{\textit{Dept. of Mechanical Engineering} \\
\textit{University of California, Berkeley} \\
reid.dye@berkeley.edu}
\and
\IEEEauthorblockN{Duncan Callaway}
\IEEEauthorblockA{\textit{Energy and Resources Group} \\
\textit{University of California, Berkeley} \\
dcal@berkeley.edu}
}

\maketitle

\begin{abstract}
Power grids are seeing more devices connected at the load level in the form of power electronics: e.g., data centers, electric vehicle chargers, and battery storage facilities. Therefore it is necessary to perform power system analyses with load models that capture these loads' behavior, which has historically not been done. To this end, we propose ZIP-E loads, a composite load model that has a ZIP load with a dynamic power electronic, or E, load model. We perform small signal and transient analysis of the IEEE WSCC 9 Bus test case with ZIP and ZIP-E load models. For small signals, we conclude that ZIP loads destabalize networks significantly faster than corresponding ZIP-E loads. In stable cases, transient results showed significantly larger oscillations for ZIP loads. Further, we find that a higher network loading condition is correlated with a higher sensitivity to load model choice. These results suggests that the constant power portion of the ZIP load has a large destabilizing effect and can generally overestimate instability, and that attention should be drawn to load model choice if operating near a stability boundary.
\end{abstract}


\section{Introduction}

This paper aims to study the impact of load modeling in power grids with high penetrations of power electronics converters. Among power systems components, loads have not seen many widely adopted models for transmission system simulation compared to generators, and transmission lines \cite{Milano_2010, machowski2020power, kundur2017power, morched1999universal, colon2023transmission}. This can be attributed to several reasons, one of which is the inherent challenge of modeling loads: power consumption is a function of many conditions, including weather, time of day, geographic location, and even human behavior. Therefore, most of the analysis done for power system stability has been using a select few model options

Data centers, electric vehicle superchargers, large-scale battery storage, and variable frequency drives (VFDs) are examples of power electronic loads connecting on the grid, and available load models do not accurately capture their dynamics for all time scales. Moreover, any analysis is limited in scope to the results produced from the choices of models. Therefore, there exists a need for power systems analyses with models that represent loads being connected on the grid. 

A survey on power system industry members in 1988  \cite{concordia_load_1982} found that the most common load modeling practice was constant current loads for active power loads and constant impedance loads for reactive power loads. ZIP loads, static linear combinations of constant impedance, constant current, and constant power loads, were also common, and only a small number of industry members used dynamic load models. 

Another survey  in 2013 \cite{Milanovic_Yamashita_Martinez_Villanueva_Djokic_Korunovic_2013} found that there was no industry standard for load models and that this choice varied significantly by continent. Variations of static ZIP loads, however, were the most common model, constituting 70\% of industry load models worldwide. Another common practice in the study was converting a ZIP load into an equivalent exponential model with a non-integer exponent. The survey also found that it is a common practice in the United States of America to use a composite load model incorporating both a static load component in the form of a ZIP load and a dynamic load component (typically an induction motor) in parallel. 

%% Other models have also been proposed in attempts to capture loads' frequency dependence, usually taking the form of a linear combination of two ZIP models.
Other load models have been proposed to capture a load's dependence on frequency. These usually take the form of a linear combination of two ZIP load models \cite{concordia_load_1982}. A similar attempt is made by the EPRI LOADSYN and ETMSP static load models \cite{concordia_load_1982}. These, however, according to the surveys, were not widely adopted by industry members.

Despite the 25 year gap between the surveys, the most recent does not show significant changes in the way industry models loads despite the increasing level of power electronics loads on the grid. This further means that corresponding analyses have not had accurate representations for power electronic loads. 

% A 2018 review on load modeling \cite{Arif_Wang_Wang_Mather_Bashualdo_Zhao_2018} provides a detailed summary of the field and introduces new models including an artificial neural network and another modeling real and reactive power consumption as a reaction to step voltages at the load. Other models rely on fitting ZIP models to real-world load data, but these are typically for individual loads rather than aggregate loads as is necessary for transmission scale analysis \cite{Zhao_Tang_Zhang_Wang_2010, Schneider_Fuller_Tuffner_Singh_2010}. %These models, however, are not based on first principles physics.

In 2022, the IEEE Standards Association and IEEE Power and Energy Society published a guide for load modeling for simulations of power systems \cite{noauthor_ieee_2022}. The guide acknowledges that only a few load models are widely accepted among industry members. % It also references more recent work that has modeled power electronic devices as loads, but these are constrained to parameter fitting of the standard static ZIP load. 
For dynamic loads, motors are the most common model \cite{agg}, and the exponential recovery load model captures aggreagate transmission-level load dynamics \cite{hill1993nonlinear}. 

Lastly, the state-of-the-art load model for phasor domain simulations is the WECC composite load model \cite{kosterev2008load} but is limited to static models for power electronic loads, except for VFDs \cite{mitra2020modeling}. It, however, does not represent inverter control loops relevant for higher frequency dynamics. A recent paper \cite{henriquez2024small} uses dynamic power electronic load models for EMT studies with line dynamics but assumes no load heterogeneity.

In sum, loads are typically modeled as static ZIP loads, with motors seldom incorporated for dynamic simulations. Further, there is no well-motivated way of choosing how a load is modeled for different kinds of studies. This suggests for the last 40+ years industry has largely been using the same models for EMT and small signal analysis.

It has been posited that power electronic loads need to be modeled as constant power loads for analysis \cite{noauthor_ieee_2022}, however, this static model choice is an approximation of a dynamic device. Therefore, while this may be an accurate modeling choice for slow time scales, we believe we should model power electronics load with corresponding dynamics for analysis on fast time scales. This is further motivated by concerns regarding current saturation limits for inverters \cite{milano2018foundations} and how this affects stability conclusions and inverter current dynamics. So, there is a gap in the literature for analysis of power systems with models that represent the new devices being connected.

Therefore, in this paper we propose ZIP-E loads, a composite load model including a static ZIP model, and a power electronic load, which we denote as an E load, for the analysis of power systems with power electronic loads. We model this load as a grid-following inverter that consumes instead of delivers power: it is a rectifier, as are the the power electronic loads described. An E load is meant to capture the dynamics of those power electronic devices, and the ZIP-E load is meant to capture heterogeneity in the load profile.

% An E load provides flexibility as it draws constant power in steady state operating conditions but also allows some slack during transients which helps by reducing the need to control both voltage and current fast enough to power a stiff constant power load.

For now, we do not incorporate motors as loads for two primary reasons: i.) we seek to identify the modeling impact of changing P loads to E loads, and ii.) most modern motor loads are interfaced with VFDs, which inherently have a power electronics interface of the form we choose to model here.

To our knowledge, this is the first small signal and EMT study of power systems with heterogeneous load models that includes physics-based models for power electronic loads.

We believe that ZIP-E loads provide a flexible alternative for load modeling during fast time scales: ZIP-E loads adopt i.) the common industry practice of ZIP loads, ii.) the notion that power electronic loads will behave as constant power loads in steady state, iii.) an energy buffer so the constant power constraint of the load is relaxed during transient events, and iv.) a physical representation of power electronic loads capturing the dynamics of new loads being incorporated into the grid. 

% While this new modeling alternative is exciting, in \cite{sig_hisk} the author supports the hypothesis that load modeling does not matter much for systems that are far from a stability boundary. We seek to test this hypothesis. This being said, 
Further, by modeling loads in this way we consider fast dynamics on both generation and load, meaning that fast line dynamics could  be relevant. We investigate their effect as well.

This research is guided by the following questions:
\begin{itemize}

    \item What, if any, qualitative differences are there between the conclusions of small signal stability analyses of power systems that use ZIP versus ZIP-E load models?
    \item Can ZIP-E loads influence inverter current dynamics in ways not otherwise manifested by ZIP models?
    
    % \item How suitable are P load models in comparison to E load models for small signal stability and short term dynamic modeling of power systems? 
    % \item Is load model choice particularly important for small signal and transient analyses?
    
    %% \item What variables are of most interest to pay close attention to under different load models?
    %% \item Are constant power load models too demanding a requirement for stability studies?

    %% \item What kind of load models are best suited to represent physical behavior for EMT studies?
    %% \item Are $E$ load models suited to represent the emerging power electronic loads on the grid?
    %% \item What load models are best suited for the different power system stability problems?
\end{itemize}

This paper's contributions are:
\begin{itemize}
    %\item \texttt{ZIPE\_loads.jl} -- A Julia-based open source modeling package for ZIP-E loads extending the modeling and simulation capabilities of \texttt{PowerSimulationsDynamics.jl} \cite{lara2023powersimulationsdynamics}.
    \item An analysis load model effects on small-signal and transient stability of the IEEE WSCC 9 Bus Test Case.
    \item A series of recommendations for load models for static and dynamic power system simulation studies.
    \item ZIP loads with larger constant power portions induce larger oscillations in transient simulations relative to ZI-E loads which are significantly more dampened and converge more regularly.
\end{itemize}

\begin{comment}
\subsection{Modeling to replicate versus modeling to gain insight}

An important distinction we would like to make is noting a difference between modeling with the objective to replicate an event that occurred in a real power system versus modeling with the objective to gain insight about a power system's behavior. 

Load modeling for replicating physical events is a challenging task because of the inherent heterogeneity in real world loads. So modeling for replication perhaps should not be the objective in general, rather the proposed objective here is modeling for qualitative understanding. 
\end{comment}

% The remainder of the paper is organized as follows. Section \ref{loads} presents the load models we use, and section \ref{load_comp} the corresponding load portfolios. Section \ref{test_case} discusses the test case we experiment with, section \ref{results} shows results and discusses analysis, and section \ref{conclusions} concludes the paper.

\section{Power System Loads}
\label{loads}

We define a load at the transmission level as an aggregation of all circuits and devices downstream from the transmission circuit, including but not limited to the distribution grid, commercial loads, industrial loads, domestic loads, and devices.

Power systems loads are categorized as static or dynamic. Static refers to the power consumption at that load being defined as an algebraic relationship between power and voltage. Dynamic refers to having a differential equation model for the variables that dictate power consumption. Composite loads combine both static and dynamic load models in parallel.

% Static load models not have any dynamics on the voltage or current at the load which dictates the power consumption at the load. They are usually represented by a set of algebraic equations. Dynamic load models are typically motor or thermostatically controlled loads and usually come in the form of a set of differential algebraic equations.

%\subsection{What is an exponential load model?}

An exponential load model, also called a voltage dependent load model, relates the power consumption of the load to a power of the normalized voltage. It takes the following form.

\begin{align}
    P_{exp} = P_0 \left( \frac{V}{V_0} \right)^{n_P}  \\
    Q_{exp} = Q_0 \left( \frac{V}{V_0} \right)^{n_Q}
\end{align}

Here, $P_0, Q_0, V_0$ are the nominal real power, reactive power, and voltage magnitude for the load. $V$ is the voltage magnitude at the load. $n_P$ and $n_Q$ are the exponents of the model.

The exponential model is inherently a static load model because the voltage magnitude $V$ which dictates the power consumed by the load is determined by the network, and does not vary as a function of the load behavior.

A ZIP or polynomial load is the convex combination of three exponential loads: a constant impedance load model (defined by $n_P = n_Q = 2$), a constant current load model (defined by $n_P = n_Q = 1$), and a constant power load model (defined by $n_P = n_Q = 0$). Mathematically it is defined as follows.

\begin{align}
    P_{ZIP} = P_0 \left(\eta_Z \left(\frac{V}{V_0}\right)^2 + \eta_I\frac{V}{V_0} + \eta_P\right) 
    \label{eq:p_zip}\\
    Q_{ZIP} = Q_0 \left(\gamma_Z \left(\frac{V}{V_0}\right)^2 + \gamma_I\frac{V}{V_0} + \gamma_P\right) 
    \label{eq:q_zip}
\end{align}

$\eta_Z, \eta_I, \eta_P$ and $\gamma_Z, \gamma_I, \gamma_P$ are weights for each load type. Electrically, it is equivalent to the three loads in parallel. %% It has been shown from data fitting processes \cite{Schneider_Fuller_Tuffner_Singh_2010,zhao_modeling_2010}, however, that ZIP loads can actually be the affine, instead of convex, combination of such loads suggesting that the overall load consumes power, but that elements of the load could be supplying power to other elements of the load.

\begin{comment}
    \begin{remark}
    In the this discussion, the power consumed by a load is determined by the voltage magnitude, $V$, at the bus which it's connected.

    For a quasi-static phasor (QSP) simulation that does not model TL dynamics, this voltage comes from the forward propagation of a power flow solution.

    For an EMT balanced $dq$ simulation as we will perform in this study, this voltage will come from the TL's capacitor's voltage state. Said capacitor will have $v_d$ and $v_q$ components and so $V = || \boldsymbol{v}_{dq} ||_2 = \sqrt{v_d^2 + v_q^2}$.
\end{remark}
\end{comment}

% \subsection{What is a power electronic load model?}

In this study, we model E loads as grid-following inverters with standard real and reactive power outer loop PI controls, PI inner loop current controls, an averaged inverter model, a phase-locked loop, and an LCL filter, as in \cite{henriquez2024small}. 

Contrary to the ZIP load, an E load is a dynamic load model because the current drawn is described by a differential equation. The power consumed by an E load model is:
\begin{align}
    P_E = \frac{1}{2}(v_d i_d + v_q i_q)\\
    Q_E = \frac{1}{2}(v_q i_d - v_d i_q)
\end{align}
$v_d, v_q$ are the load bus' per unit  capacitor voltage $dq$ components. $i_d$ and $i_q$ are the per unit current $dq$ components drawn by the inverter defined by a system of ordinary differential equations. A detailed modeled can be found in \cite{henriquez2020grid}.

% \subsection{What is a ZIP-E load model?}

A ZIP-E load is a convex combination of a ZIP load and an E load model, and the power it consumes is as follows. 
% \begin{align}
%     P_{ZIP-E} &= P_{ZIP} + P_E \\ & = P_0\left(\eta_Z\left(\frac{V}{V_0}\right)^2 + \eta_I\frac{V}{V_0} + \eta_P +  \frac{1}{2} \eta_E \left(v_d i_d + v_q i_q\right)\right) \\
%     Q_{ZIP-E} &= Q_{ZIP} + Q_E \\ &= Q_0\left(\gamma_Z\left(\frac{V}{V_0}\right)^2 + \gamma_I\frac{V}{V_0} + \gamma_P +  \frac{1}{2} \gamma_E \left(v_q i_d - v_d i_q\right)\right)
% \end{align}
%
% this fits better width-wise and makes for a simpler final expression
% though I do have a tendency to over-vectorize all notation
% 
% bold L was just a random choice. feel free to 
% rename to something better
%
\begin{align}
P_{ZIP-E} &= P_{ZIP} + P_E  
= P_{0} \left( \boldsymbol{\eta}^{\top}\boldsymbol{L_P} \right) \\
Q_{ZIP-E} &= Q_{ZIP} + Q_E 
= Q_{0}\left( \boldsymbol{\gamma}^{\top}\boldsymbol{L_Q} \right)
\end{align}
where 
\begin{align}
\boldsymbol{\eta}&=\begin{bmatrix}
\eta_{Z} &
\eta_{I} &
\eta_{P} &
\eta_{E}
\end{bmatrix}^{\top}
\\
\boldsymbol{\gamma}&=\begin{bmatrix}
\gamma_{Z} &
\gamma_{I} &
\gamma_{P} &
\gamma_{E}
\end{bmatrix}^{\top}
\end{align}
and
\begin{align}
\boldsymbol{L_P}&=\begin{bmatrix}
    \left( \frac{V}{V_{0}} \right)^{2}  &
    \left( \frac{V}{V_{0}} \right)^{1} & % makes it clearer that
    \left( \frac{V}{V_{0}} \right)^{0} & % these are just decreasing exponents
    \frac{1}{2}(v_{d}i_{d}+v_{q}i_{q})
    \end{bmatrix} ^{\top}
\\ 
\boldsymbol{L_Q}&=\begin{bmatrix}
    \left( \frac{V}{V_{0}} \right)^{2}  &
    \left( \frac{V}{V_{0}} \right)^{1} & % makes it clearer that
    \left( \frac{V}{V_{0}} \right)^{0} & % these are just decreasing exponents
    \frac{1}{2}(v_{q}i_{d}-v_{d}i_{q})
    \end{bmatrix}^{\top}
\end{align}

Note the addition of the $\eta_E$ and $\gamma_E$ terms which capture the power consumed by the E part of the load. Moreover, it is now the case that the load consumes real and reactive power not only as a function of the voltage at the load bus, but also as a function of the current drawn by the load.
\footnote{We develop \href{https://github.com/gecolonr/ZIPE_loads.jl}{\texttt{ZIP-E\_loads.jl}}, an open source Julia-based package to extend the modeling capabilities of \texttt{PowerSimulationsDynamics.jl (PSID.jl)}. It leverages already available models in \texttt{PSID.jl} to allow the user to model ZIP-E load variations.}

\section{Load composition}
\label{load_comp}

Based on a study of the literature, there is no standardized way to choose ZIP load coefficients based on an understanding of the load behavior. Therefore, we outline modeling options based on what industry has done and propose alternatives.

One option is as it was done in the 1980s \cite{concordia_load_1982}: real power loads as static constant current loads, and reactive power loads as static constant impedance loads. Another is what was done in the 2000s: real and reactive power loads as static ZIP loads, as in \eqref{eq:p_zip} and \eqref{eq:q_zip}. However, it's not clear how to select coefficients \cite{Milanovic_Yamashita_Martinez_Villanueva_Djokic_Korunovic_2013}. Based on \cite{Milanovic_Yamashita_Martinez_Villanueva_Djokic_Korunovic_2013}, the USA is the only country where a substantial number of industry members run simulations with composite load models. \cite{Milano_2010} proposes a few coefficient options for the last two options.

\begin{comment}
    This is typically done in the following form.

    \begin{align}
        S_{tot} = \nu S_{ZIP} + (1-\nu)S_{IM}
    \end{align}
    
    Here, $S_{ZIP} = P_{ZIP} + j Q_{ZIP}$ and $S_{IM} = P_{IM} + j Q_{IM}$, where $P_{IM}$ and $Q_{IM}$ follow dynamics of a typical third order induction motor model \cite{noauthor_ieee_2022}. $\nu$ is a coefficient to determine what percent of the load is static versus dynamic.
\end{comment}

\subsection{New scenarios and benchmarks}

% We consider a constant impedance test case as a benchmark since this is the most common load type in research. We consider constant power too as a benchmark, and we also consider an E load for comparison. We also consider cases of fixed ZI ratio, while changing P or E, that is, ZIP loads and ZI-E loads, for further comparison. We consider low, medium, and high E load test cases, while keeping the ratio of the remaining load types fixed. We include these in Table \ref{table:load cases}, where $x,y,z \in [0,1]$ such that the convex combination follows.

We consider a constant impedance test case as a benchmark, as this is the most common load type in research. We consider cases with P or E load contributions (through $\eta_P$ and $\eta_E$, respectively) in 10 percentage point increments from $0$ to $100\%$ with the remainder of the load constituting of equal parts Z and I, thus comparing ZIP loads to ZI-E loads by choosing $\boldsymbol{\eta}, \boldsymbol{\gamma}$ as follows. 
$$
\boldsymbol{\eta} = \boldsymbol{\gamma} =\begin{bmatrix}
    \frac{1-x}{2} \\
    \frac{1-x}{2} \\ 
    x \\
    0
\end{bmatrix}, \begin{bmatrix}
    \frac{1-x}{2} \\
    \frac{1-x}{2} \\ 
    0 \\
    x
\end{bmatrix}\quad \forall x\in \{0, 0.1, \dots, 1.0\} 
$$

Note that for each $x$ we have a particular load portfolio. Constant power and full E loads are defined by $x = 1.0$. This comparison will allow us to test differences of the effects modeling power electronic loads as P loads or E loads.
% needs to include info about the other set of parameters used. Or we can just leave the plot legend to show that and remove this table.
% \begin{table}[]
%     \begin{center}
%     \caption{Load portfolios}
%     \begin{tabular}{|| c | c | c | c | c ||} 
%      \hline
%      Case & $\eta_Z = \gamma_Z$ & $\eta_I = \gamma_I$ & $\eta_P = \gamma_P$ & $\eta_E = \gamma_E$ \\ [0.5ex] \hline\hline
%      Constant Impedance & 1.0 & 0.0 & 0.0 & 0.0 \\ 
%      \hline
%      Constant Power & 0.0 & 0.0 & 1.0 & 0.0 \\
%      \hline
%      Full E load & 0.0 & 0.0 & 0.0 & 1.0 \\
%      \hline
%      ZIP loads & $x$ & $x$ & $1-2x$ & 0.0 \\
%     \hline
%      ZI-E loads & $x$ & $x$ & 0.0 & $1-2x$ \\
%      \hline
%      ZIP-E loads & $x$ & $y$ & $z$ & $1-x-y-z$ \\
%      %\hline
%      %Low E load & 0.3 & 0.3 & 0.3 & 0.1 \\
%      %\hline
%      %Medium E load & 0.2 & 0.2 & 0.2 & 0.4 \\ 
%      %\hline
%      %High E load & 0.1 & 0.1 & 0.1 & 0.7 \\[1ex] 
%      %\hline
%      %Low P High E load & 0.15 & 0.15 & 0.15 & 0.55 \\[1ex] 
%      %\hline
%      %High P Low E load & 0.15 & 0.15 & 0.55 & 0.15 \\[1ex] 
%      \hline
%     \end{tabular}
%     \title{Load cases}
%     \label{table:load cases}
%     \end{center}
% \end{table}

\section{Test Case}
\label{test_case}

\begin{figure}
    \centering   \includegraphics[scale=0.5]{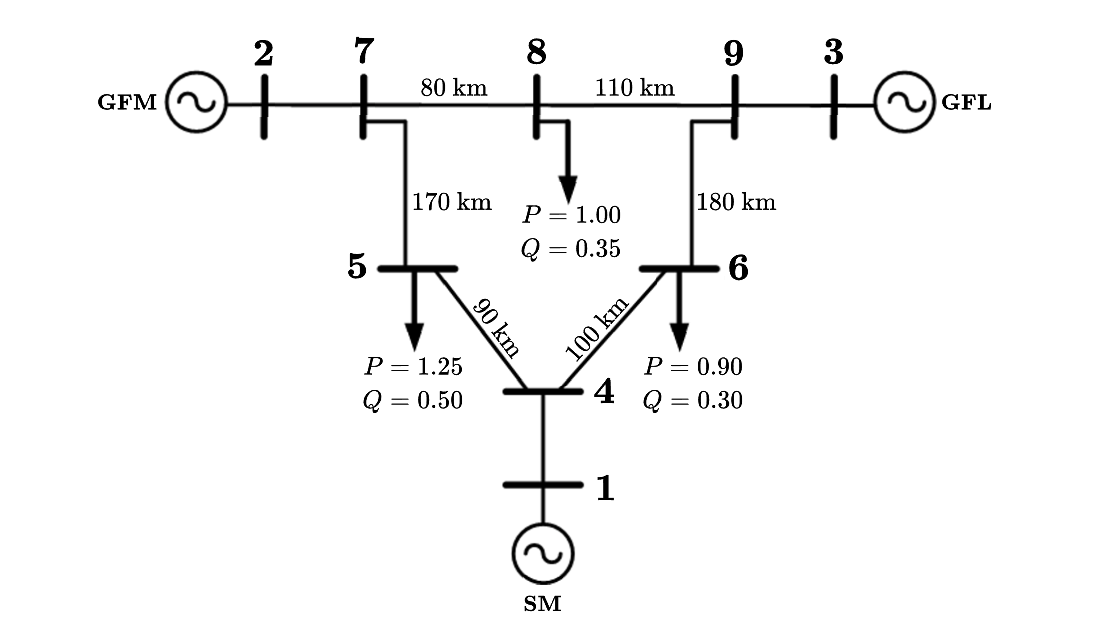}
    \caption{Modified IEEE 9 bus test system. SM at bus 1 (reference bus), VSM GFM at bus 2, GFL at bus 3. Note that we do not use a slack bus for our simulations.}
    \label{fig:9bus_schematic}
\end{figure}

In this work, we use the IEEE WSCC 9 Bus test case shown in Fig. \ref{fig:9bus_schematic}. It is composed of a synchronous machine (SM), a grid-forming converter (GFM) operating as a virtual-synchronous machine (VSM), and a grid following inverter (GFL) as sources. It has three loads, and six transmission lines. For lines, we use algebraic and dynamic $\pi$ line models, $statpi$ and $dynpi$ respectively. Models and data for these devices are according to \cite{colon2023transmission}. We stay consistent in our choice of model for purposes of comparing results across papers.

\subsection{Experiments}

The network's nominal operating condition is found by solving a power flow, and then loads are scaled by a constant factor we denote $load \ scale$. We then scale all generator power set points in proportion to $load \ scale$. For each load model, for each line model, and for each $load \ scale$ value, we run a small signal and transient analysis. 

For the small signal studies, given a single $load \ scale$ value (or, equivalently, a network loading condition), all line and load models will produce the same power flow solution. We verify the stability of this operating condition to address our guiding questions.

For the transient analysis, we trip the heaviest loaded branch as given by the power flow solution, which is the line connecting buses 4 and 5 in this case. We choose to measure the current flowing into the network at Bus 3 as a proxy variable for inverters since it could be saturated if control limits are reached. We evaluate time domain simulations to see current behavior difference to address our guiding questions.

\begin{figure*}[htbp]
    \centering
    % \includegraphics[width=\textwidth,scale=0.3,trim=0 3cm 0 4cm,clip]{figures/eigplot2_300dpi.png}
    % \vspace{-0.5cm}

    \includegraphics[width=\textwidth,scale=0.3,clip]{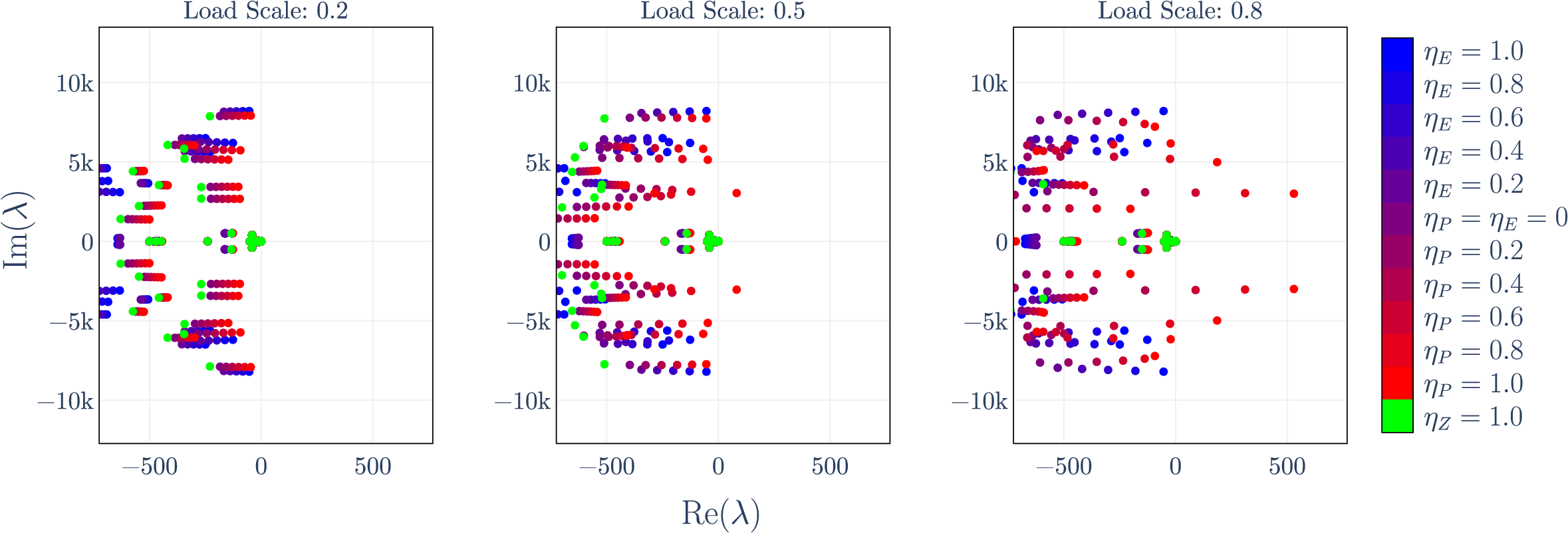}
    \vspace{-0.5cm}
    
    \caption{Network eigenvalues with $dynpi$ line model at $load\ scales$ of 0.2, 0.5, and 0.8 increasing to the right.}
    \label{fig:eigenvalue_results}
\end{figure*}

\begin{figure}[htbp]
    \centering
    % \hspace*{-0.5cm}\includegraphics[scale=0.3,trim=0.7cm 4.7cm 0 5cm,clip]{figures/eigplot1_300dpi.png}
    \hspace{-0.5cm}\includegraphics[scale=0.3, clip]{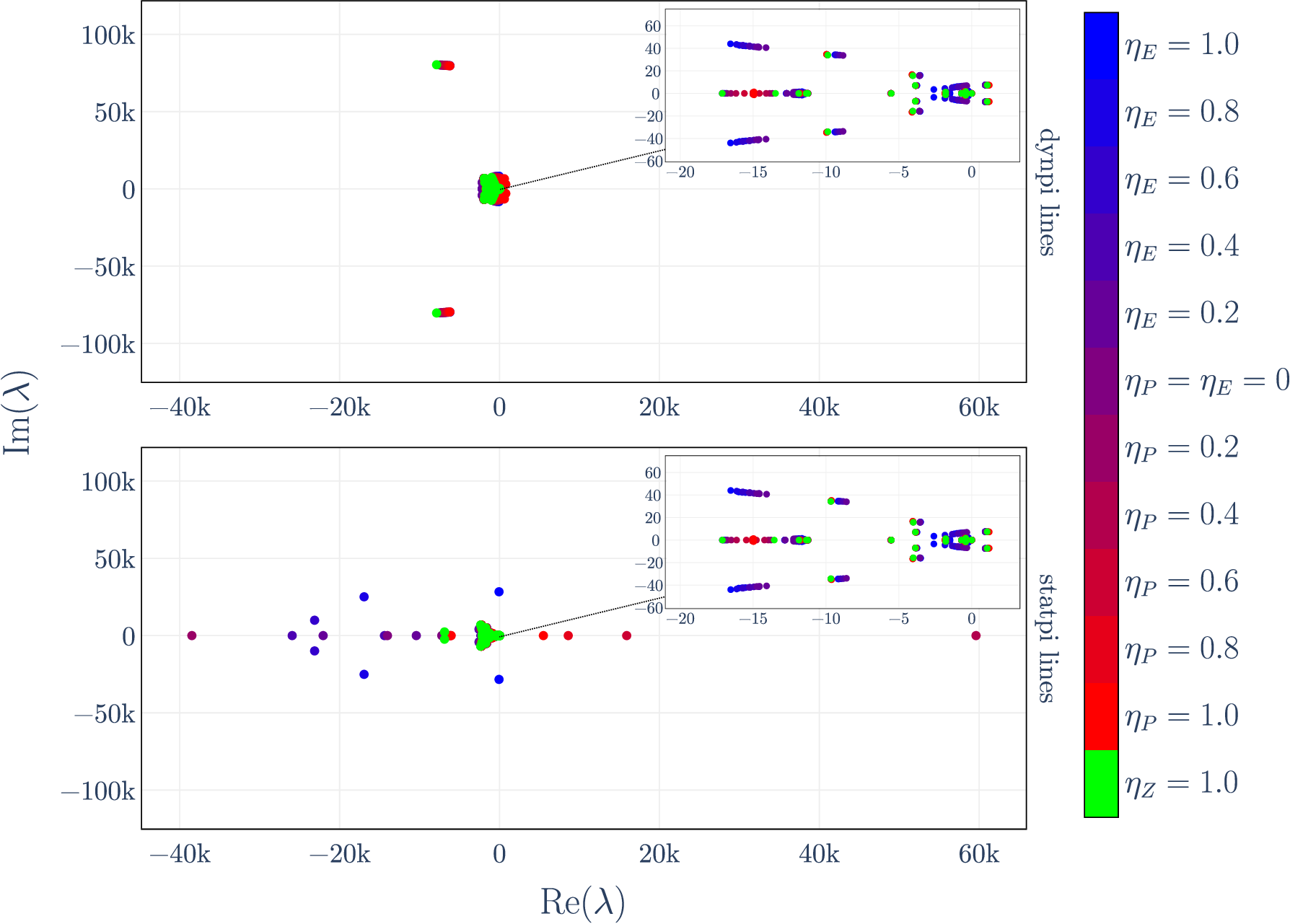}
    % \includesvg[scale=0.2]{figures/transient_current_plot.svg}
    % \vspace{-0.5cm}
    \caption{Eigenvalues at $load\ scale=1.0$ for $dynpi$ (top) and $statpi$ (bottom) line models. ZI-E loads are in blue  and ZIP loads are red in 10\% increments as given by the heatmap bar on the right.}
    \label{fig:eigenvalues_withlines}
\end{figure}

\section{Simulation Results and Analysis}
\label{results}

Code and results are publicly available in our \href{https://github.com/gecolonr/loads}{GitHub repository} for reproduction, which includes a set of interactive plots for the reader to further explore our results if interested.

%% Added all the plots here. Please rearrange to your liking.
%% I also think i'll increase the text size on all of them tomorrow; it's kinda small

\begin{figure*}[htbp]
    \centering
    % \includegraphics[width=\textwidth,scale=0.3,trim=0 3cm 0 3.5cm,clip]{figures/transient2_300dpi.png}
    % \vspace{-0.5cm}

    \includegraphics[width=\textwidth,scale=0.3,clip]{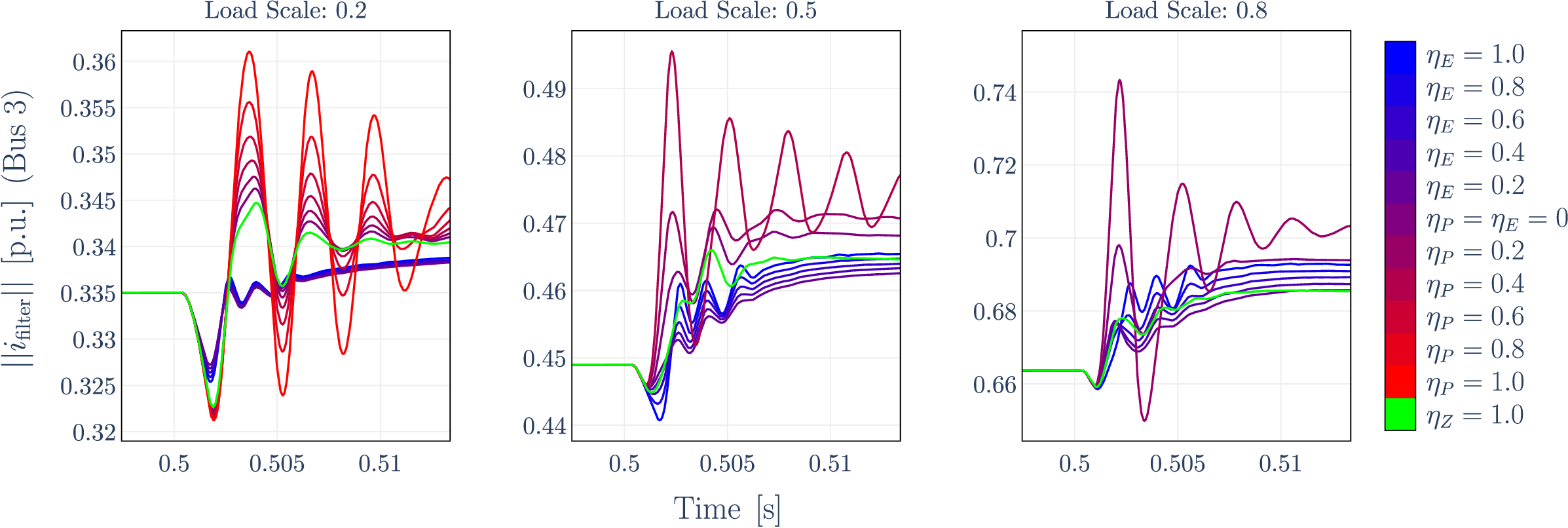}
    \vspace{-0.5cm}
    \caption{Bus 3 inverter current magnitude after a branch trip on line 4-5 with $dynpi$ lines.}
    \label{fig:transient_results}
\end{figure*}

\begin{figure}[htbp]
    \centering
    % \hspace*{-0.5cm}\includegraphics[scale=0.3,trim=0.5cm 4.5cm 0 5cm,clip]{figures/transient1_300dpi.png}
    % \vspace{-0.5cm}
    \hspace{-0.5cm}\includegraphics[scale=0.3,clip]{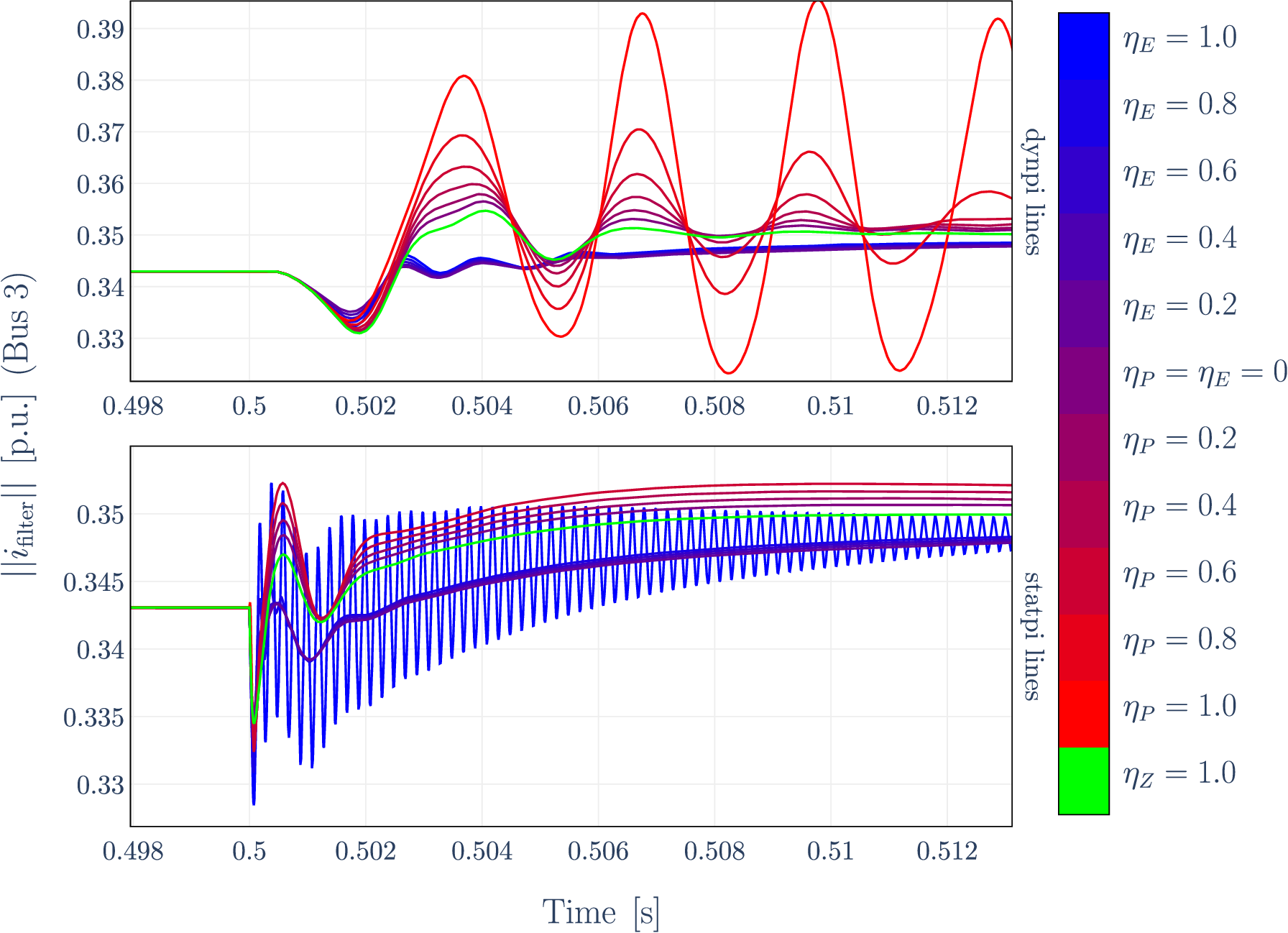}
    \caption{Transient simulation of Bus 3 inverter current magnitude after a branch trip of line connecting Bus 4 and Bus 5. $load\ scale = 0.25$.}
    \label{fig:transient_withlines}
\end{figure}

\subsection{Small Signal Analysis}

\subsubsection{Load model effect}
When comparing eigenvalues of ZI-E loads with high E values to ZI-E loads with low E values, there was no consistent trend. Some shifted to the right, others to the left, this is likely dependent on the influence of particular states on those eigenvalues. This is also true for ZIP loads. This can be appreciated in Fig. \ref{fig:eigenvalue_results}.

Results further showed that at low network loading levels, load model variations had limited effect on eigenvalues and thus on stability conclusions. This can be seen in the leftmost plot in Fig. \ref{fig:eigenvalue_results}, with all load models concluding in a stable operating condition. This suggests that when the network has enough buffer between the operating condition and the instability boundary, the choice of load model has a small impact on system behavior and stability conclusions.

As we increase the network loading, however, we find that ZIP load cases move the eigenvalues to the right more quickly than ZI-E loads. Several cases showed a stable operating conclusion with ZI-E loads, but an unstable conclusion with ZIP loads. The center plot in Fig. \ref{fig:eigenvalue_results} shows an example.

As the network loading further increased, we found that the eigenvalues for ZIP load cases were very far right on the complex plane, whereas eigenvalues for ZI-E cases were unstable but very close to the $j \omega$ axis with real parts less than 1. This suggests that, despite being an unstable operating condition, it is much more stable relative to the common industry practice of ZIP loads. This can be appreciated on the rightmost plot of Fig. \ref{fig:eigenvalue_results}. Further, because these ZI-E cases are barely unstable, this operating condition might be stable with differently tuned controller gains for the generating devices.

Lastly, in addition to eigenvalues moving to the right under higher loading, the effect of load model on a particular eigenvalue is more pronounced as evidenced by more spread eigenvalue maps on Fig. \ref{fig:eigenvalue_results} as loading increases.

\begin{comment}
    
At higher frequencies ZIP loads relative to the same portfolio of ZI-E loads tended to be more to the left of the complex planes surprisingly suggesting more damping.

Looking at eigenvalues relatively close to the origin is a good way to isolate the effect of load models since it's quite clear that line models don't significantly influence the least stable modes.

At low frequencies, less than 0.15 Hz, interestingly ZIP loads eigenvalues seem to be slightly more to the left on the complex plan, potentially indicating more stability. Intuitively, this is not that incredibly surprising actually in relation to our argument of constant power loads being suitable for steady state analysis, to which 0.15 Hz is relatively close to.

\end{comment}

These results support the hypothesis that a low stability margin correlates with higher sensitivity to load model choice presented in \cite{sig_hisk}. In situations where the distance to the instability boundary is small, modeling constant power loads with the typical algebraic model could lead to significantly overestimated instability. The E component seems to aid in remedying this.

\subsubsection{Load model effect with line dynamics assumption}
A consistent theme across all cases was the stabilizing effect of the $dynpi$ line model, particularly at higher load scales. An example of this can be seen in Fig. \ref{fig:eigenvalues_withlines} where eigenvalues for the $stapi$ model have significantly larger real parts relative to $dynpi$ model. For the $statpi$ model we found several cases in which there is a singularity where an eigenvalue moves from minus infinity to positive infinity as loading is increased, suggesting a modeling limitation. This does not happen for the $dynpi$ model. Further, $dynpi$ lines tended to filter out many high frequency eigenvalues with the exception of complex conjugate pairs with frequency of about 12.5 kHz suggesting they could be high frequency states of the dynamic lines, these are the cluster on the top plot of Fig. \ref{fig:eigenvalues_withlines}. %A participation factor analysis shows that inverter currents states do not contribute to these modes. 
All other eigenvalues had frequencies less than 1.2 kHz, whereas for the $statpi$ case, many unfiltered modes had frequencies of about 4-5 kHz. Further, we had several line parameters and changing between them didn't significantly affect the placement of the least stable eigenvalues nor the dynamic behavior of the variables we considered. 

\subsection{Transient analysis}

For the transient analysis, we measure the magnitude of the GFL inverter's filter current at Bus 3.

\subsubsection{Load model effect}
We typically saw two families of traces post disturbance: one trace family for ZIP loads and another for ZI-E loads. When they both converged, they had the same steady-state solution but different ways to reach it. Cases with ZIP loads had larger overshoots relative to cases with ZI-E loads. 

Further, ZIP cases with larger P percentages tended to have larger amplitude than those with lower P percentages. The same was true for ZI-E cases. 
%Under low loading conditions, ZIP loads cases with $\eta_P > 0.6 $ or higher were generally unstable, but all ZI-E cases, were stable, even with an E percent of 100 \%.
This can be appreciated in the leftmost plot of Fig. \ref{fig:transient_results}. Further, as $load\ scale$ increased, cases with higher ZIP were the first to destabilize, which is consistent with the small signal analysis. This is further appreciated in the center and rightmost plots in Fig. \ref{fig:transient_results} where several of the high-P cases fail to converge, and those that do converge have significantly larger overshoots relative to all the ZI-E cases. This implies that E loads induced more damping relative to P loads for both low and high-frequency oscillations.

% From a numerical stability perspective, the susceptibility to high-amplitude, high-frequency oscillations made the ZIP loads numerically very unstable, especially with $statpi$ lines. When ZIP loads did converge on $statpi$ lines, there was immense variation in results from slight variation in parameters, suggesting limited use. In contrast, E loads were much more numerically stable, producing results that matched the general behavior of the ZIP loads but without the high-frequency components.
% From a numerical stability perspective, the very high stiffness and high-amplitude, high-frequency oscillations made ZIP loads both unstable and expensive to compute, especially with $statpi$ lines. Dynamics in the loads and lines tended to remove stiffness on small scales while preserving long-term behavior. One potential explanation for this is that stiff static elements can push inverters to extreme states due to their fast reaction time, in contrast to a synchronous machine, whose large rotational inertia helps it add limited slack. 

The long-term oscillations of the ZI-E loads were not ultimately problematic. In stable cases, they disappear quickly, and in unstable cases, they follow a similar trend to ZIP loads. 

% One notable caveat to this analysis is the Full E load, which exhibited significant constant high-frequency oscillations. It seems that pure power-based loading is flawed, and the ZI component has some damping effect on the inverter in the E load. As it is already accepted that constant power loads destabilize the grid, it seems reasonable to conclude that full power-based loads should be avoided.

\subsubsection{Load model effect with line dynamics assumption}

High-frequency oscillations on $statpi$ lines were seen for both high-E and high-P loads, but decreased rapidly with decreasing E or P percentage. Only high-E loads show these oscillations because in high P cases they grew too large and the simulations did not converge. % Smaller oscillations are also visible for $\eta_E=0.8$, but they are damped within the first millisecond. This suggests a $\eta_E$ or $\eta_P$ threshold for stability in this mode. 
In most cases, the $dynpi$ line model clearly filters to 4-5 kHz frequency modes in the system present in the $statpi$ case. This is consistent with the small signal analysis, and can be appreciated in Fig. \ref{fig:transient_withlines}.

\subsection{Network loading}

Under this SM-GFM-GFL generation configuration, the network loading needed to be low ($load\ scale = 0.2$) to ensure a stable operating condition for all load and line models. We compared this to the case with SMs in all generation buses and found that for $load\ scale = 1.0$ 
the network equilibrium was stable except for ZIP loads with $\eta_P > 0.4$ with both $stapi$ and $dynpi$ lines. This complements the finding that ZIP loads are the most difficult loads to stabilize, and speaks to the relevance of generation configuration contributing to the stability of the equilibrium condition. % Further, it implies that the loading a network can tolerate is low without inducing an unstable operating condition when there are power electronics both at the generation and load levels.

\subsection{Computational burden}

Networks with ZI-E loads have about 30 more dynamic states than those with ZIP loads due to the three E loads. Despite that, we found that $statpi$ cases' runtime was not significantly impacted, and $dynpi$ simulations with ZI-E loads ran $7-10 \times$ faster than corresponding ZIP load cases, suggesting that ZIP loads can be more computationally intensive than corresponding ZI-E loads. 

\section{Conclusions}
\label{conclusions}

In this work we perform small signal and transient stability analyses for power systems using a new composite load model, ZIP-E loads, for more realistic power electronics load modeling of power systems at the transmission level. We further develop \texttt{ZIPE\_loads.jl}, an open-source load modeling package to model these loads in \texttt{PSID.jl}.
We perform experiments on the IEEE WSCC 9 Bus test case to test how changing line and load models affect stability. 

In the context of our guiding questions, we found that: i.) The choice of load model matters less when the network is lightly loaded, which corresponds to when the operating condition is far from its stability boundary. As the system is more heavily loaded and approaches the stability boundary, eigenvalue placement changes more drastically in relation to the load model, and ZIP loads result in an unstable operating condition significantly earlier than corresponding ZI-E loads. ii.) ZI-E loads showed a different family of traces relative to ZIP loads: they showed more damping, significantly smaller overshoots, and convergence to the slow modes much quicker. While overall trajectories of ZI-E loads are different than ZIP loads, they have the same steady state behavior in cases where both converge, as expected.

We conclude that in general load models are important for both small signal and transient analysis for power system, especially under stressed conditions. Approximating power electronics loads behavior as ZIP loads could result in unrealistic unstable conclusions far more quickly than could actually happen. The conclusions drawn from cases with ZI-E loads can be significantly different from those with ZIP loads both for small signal and transient analysis. Because of the large amounts of power electronic loads getting connected on the network, we believe ZIP-E loads make a promising attempt and capturing their dynamics so that we can arrive at reliable and trustworthy results.

This work results in the following recommendations: i.) if a particular system is operating at a condition far from a stability boundary, and that is known prior to running a simulation, pick the least expensive computational load or easility to interpret model, a constant impedance model would be suitable,
ii.) if the network is close to a stability boundary, constant power loads will overestimate instability, therefore, use ZI-E load models, iii.) between $statpi$ and $dynpi$ line models, use $dynpi$ for both transient and small signal analyses.

Future work includes changing the location of the generators, the generation portfolio, and varying the system's grid strength to explore low-frequency oscillations.

\section*{Acknowledgements}
This work was possible thanks to the partial funding of PSERC grant \# from \textit{Réseau de Transport d'Électricité, France}. In particular thanks to A. Guironnet, M. Chiaramello, P. Panciatici for fruitful discussions.

\bibliographystyle{IEEEtran}
\bibliography{refs}

\end{document}